# Electrical and optical properties of ITO thin films prepared by DC magnetron sputtering for low-emitting coatings

Hadi Askari[1*], Hamidreza Fallah[1], Mehdi Askari[2], Mehdi Charkhchi Mohmmadieyh[1]

[1] Faculty of Science, Department of Physics, Isfahan University, Isfahan, Iran

[2] Department of electrical engineering, Behbahan Khatam Alanbia University of Technology, Behbahan, Iran

*corresponding author: hadi.askari1981@gmail.com

**Abstract:** Optimized DC magnetron sputtering system for the deposition of transparent conductive oxides (TCOs), such indium tin oxide (ITO) on glass substrate has been applied in order to achieve low-emitting (low-e) transparent coatings. To obtain the concerned electrical resistance and high infrared reflection, first the effect of applied sputtering power then oxygen flow on the properties of films have been investigated. The other depositions parameters are kept constant. Film deposition at at temperature 400 ºC in oxygen flow of 3sccm results in transparent and infrared reflecting coatings. Under this condition the highest attained average reflectance in the infrared is ($\lambda$=3-25 μm) 89.5% (lowest emittance equals to less than 0.11), whereas transparency in the visible is 85% approximately. Plasma wavelength and carrier concentration was measured

## Introduction

Thin films of Indium Tin Oxide (ITO) have been extensively used in numerous electronic application such as antistatic applications [1], architectural coatings (low emissivity glazing, solar control and antireflective coatings), [2-5] transparent electrodes in solar cells and flat panel display [6], OLEDs [7], due to their unique characteristics such as high conduction, high optical transmittance in visible area, high infrared reflectance and excellent substrate adhesion [3, 8]. Electrical and optical properties of ITO films are sensitive to production conditions. There are several different methods for deposition ITO thin films such as DC and RF magnetron sputtering [6, 9], electron beam evaporation [8], pulsed laser deposition [7]. with a high sputtering rate and good films performance, DC-magnetron sputtering is used widely [10], we deposited ITO thin films on large scale glass and investigate its properties in order to fabricate Low-E coatings which transmit visible wavelengths and are good reflectors for infrared waves, especially far infrared waves [3, 4]. In fact, we measure electrical and optical properties of ITO thin films, which have been deposited on glass, using DC magnetron sputtering as a function of oxygen flow and applied power and then we investigated the effect of these parameters on the properties of prepared films.

Researches and studies show that ITO films must have a sheet resistant less than 15Ω/sq to obtain IR reflectance over 80%; In this case, transmission in visible area will be more than 80%. Also according to available reports[8], carrier concentration on these films is between $4-8 \times 10^{20} cm^{-3}$ and in such case; reflection in IR area reaches almost near 50% (λ=0.7-3µm) and average reflection in IR area reaches about 90%. Low emitting coatings are in fact those coatings which are capable to save energy consumption. Low emission glasses which are called Low-E are those glasses which are coated with a thin film and therefore their heat exchanges decrease. This coating reflects heat waves while allows visible light transmission.

**Experiment**

A sputtering system was used to deposit the ITO films by the DC-magnetron sputtering method. The target dimensions were $130 \times 20 cm^2$. The target used for these evaluations was a ceramic target containing 10wt% $SnO_2$ and 90wt% $In_2O_3$. Float glass was used as substrate. The substrates were cleaned in an ultrasonic cleaner for 10min with acetone. The deposition was carried out an argon atmosphere. Argon and oxygen gas flows were controlled by mass flow controllers.

At first in this study, the influence of increasing applied sputtering power was investigated to achieve a sheet resistance lower than 30Ω/sq, then the effect of oxygen content on the films. Deposition was done in 400 ºC. All other deposition parameters are kept constant.

The sheet resistance was determined with four point probe system FPP5000 supplied by Veeco. Transmission and reflection measurements were performed with UV-3100-Shimadzu UV-VIS-NIR and Far-IR spectrophotometer. The plasma wavelength is defined at T=R where the dielectric-like transmission equals the metallic-like IR reflection [7]. Free carrier concentration was calculated from T and R based on Drude theory. The plasma resonance frequency ($\omega_p$) is given by

$$\omega_p^2 = (Ne^2)/\varepsilon_o \varepsilon_\infty m_e^* \quad [7] \tag{1}$$

## Results and Discussion

### Electrical Properties

Electrical properties of ITO films depend on film combination and deposition parameters such as applied sputtering power, oxygen flow, substrate deposition temperature, etc. Table 1 shows electrical and optical properties of deposited films in 400°C and 6.1sccm oxygen flow considering the changes of sputtering power.

It can be seen, at power 4.3KW, sheet resistance is minimized. Also carrier concentration has increased at first and then it has almost remained constant. Decrease of initial resistance may be due to increase of carrier concentration.

The oxygen flow was also found to affect the electrical properties of the ITO films. Table 2 illustrates the variation of sheet resistance as a function of oxygen flow in the sputtering atmosphere for ITO films grown at a deposition temperature 400 °C and sputtering power applied 4.3KW. The resistance of ITO films decreases with decreasing oxygen from 6.1 to 5sccm due to an increase in the number vacancies. The oxygen vacancies create free electrons in the films because one oxygen vacancy creates two extra electrons. The increase in the number of oxygen vacancies leads to an increase in carrier density and a consequent decrease in resistance. However, the resistance of ITO films increased with the further decrease in oxygen flow (<5sccm). This increase in resistance may be due to the fact that severe oxygen deficiencies may deteriorate the crystalline properties (see Fig. 1).

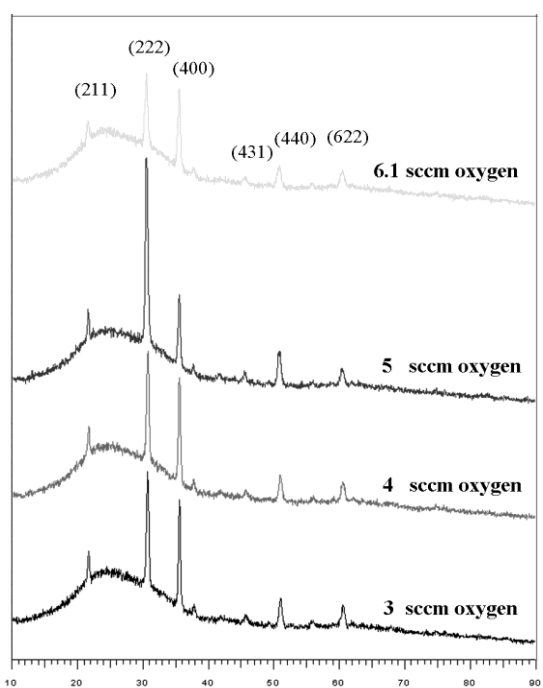

Figure 1: X-ray diffraction patterns for the ITO films grown on glass at different oxygen flow. Substrate deposition temperature was 400°C.

The existence of minimum resistivity is a well known behavior of ITO [6, 11]. A low sheet resistance has been obtained by changing the sputtering power and oxygen flow which can be used as Low-E coatings for energy efficient windows.

**Optical properties**

Optical properties such as transmission and reflection have been measured by using two spectrophotometers which worked in UV-VIS-NIR and FAR-IR region, respectively.

In near IR region, the interaction of free electrons with incident radiation occurs due to the high number of free electrons in the material. This interaction may lead to polarization of the radiation within the material and thus, affect the relative permittivity $\varepsilon$. This optical phenomenon can be understood based on Drude's theory. According to this model, $\varepsilon^*$ can be written as [2, 6-8]

$$\varepsilon^* = (n - ik)^2 = \varepsilon_r + i\varepsilon_i = \varepsilon_\infty - \frac{Ne^2}{\varepsilon_\circ m^*(\omega^2 - i\omega/\langle\tau\rangle)} \quad (2)$$

Where $\varepsilon_\infty$ is the high frequency dielectric constant, $m^*$ is the electron effective mass and $\langle\tau\rangle$ is the averaged relaxation time. $n$ and $k$ are the optical constants which determine the reflectance and absorptance spectra of the film.

Applying this result to transparent conductors with high carrier concentrations about $(10^{19} - 10^{21} cm^{-3})$ in the short wavelength approximation (i.e. $\omega\langle\tau\rangle \gg 1$) which holds well for $SnO_2$ and $In_2O_3$ we can write the expression for the real part of $\varepsilon$

$$\varepsilon_r = \varepsilon_\infty - \frac{Ne^2}{\varepsilon_\circ m^* \omega^2} \quad (3)$$

Under the condition $\varepsilon_r = 0$ eqn. (3) yields the plasma wavelength

$$\lambda_p = 2\pi c \left(\frac{\varepsilon_\infty \varepsilon_\circ m^*}{Ne^2}\right)^{1/2} \quad (4)$$

Where $e = 1.6 \times 10^{-19} C$, $\varepsilon_\circ = 8.85 \times 10^{-12} As/Vm$ and $\varepsilon_\infty = 4$ represent the dielectric constants of the medium and free space, respectively, $m_e^* = 0.4 m_e (m_e = 9.11 \times 10^{-31} Kg)$ is the effective mass of charge carrier, and $N$ is the carrier concentration [12, 13]. $\gamma$ is equal to $1/\tau$, where $\tau$ is the relaxation time, which is assumed to be independent of frequency and is related to mobility.

The absorption peak in Fig. 2 is due to the plasma resonance, hence the cutoff wavelength (or plasma wavelength) is obtained considering optical data taken from spectrophotometer in the manner that $\lambda_p$ is the wavelength which is T=R where the dielectric-like visible transmission equals the metallic-like IR reflectance [7].

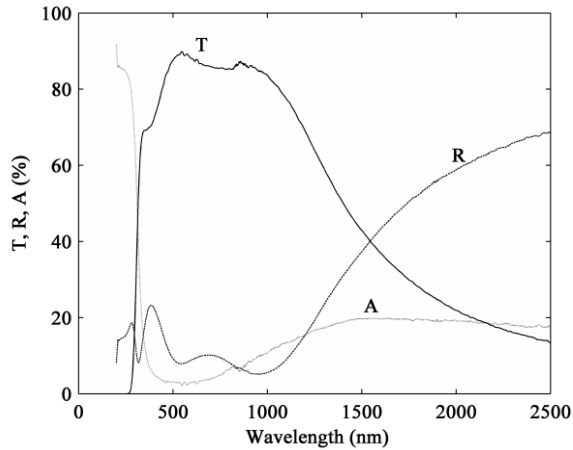

Figure 2 transmission T, Reflectance R and Absorption A for the ITO films grown at 400ºC and 4 Sccm Oxygen flow.

In tables 1 and 2, plasma wavelength changes have come as a function of sputtering power applied on target and oxygen flow. The observed shift in the plasma resonance wavelength with N for ITO films is in agreement with eqn. (3) (see Fig. 2).

Optical transmission properties of films have been considered by sputtering power changes applied on oxygen flow. Table 1 shows the effect of applied sputtering power on reflection of those films deposited in 400°C. It is clear from reflection that its amount has considerably increased by increase of power and according to Drude theory it is due to increase of carrier concentration which causes plasma wavelength to move towards shorter wavelengths, but sample No. 3 shows amount of reflection almost as same as sample No.2, therefore the effect of oxygen flow will investigate on sample No.2.

**Table 1** Variation of electrical an optical properties for the ITO films grown at different sputtering power. All films were deposited in 6.1 Sccm oxygen flow and the substrate deposition temperature was 400ºC

| Sample No. | Power (KW) | Sheet resistance (Ω/sq) | Carrier density (×10²⁰cm⁻³) | Plasma wavelength (nm) | Average transmission ($\lambda$ =0.4 – 0.7μm) | Average reflection ($\lambda$ =0.7 – 2.5μm) | Film thickness (nm) |
|---|---|---|---|---|---|---|---|
| 1 | 3.75 | 23 | 3.76 | 2180 | 84.2 | 18.5 | 94 |
| 2 | 4.3 | 12.6 | 6.36 | 1678 | 86.5 | 34.4 | 170 |
| 3 | 5 | 13.0 | 6.40 | 1670 | 86.2 | 34.5 | 172 |

Table 2 Variation of electrical an optical properties for the ITO films grown at different oxygen flow. All films were deposited in 4.3KWsputtering power and the substrate deposition temperature was 400ºC

| Sample No. | Oxygen (Sccm) | Sheet resistance ($\Omega$/sq) | Carrier density ($\times 10^{20}$cm$^{-3}$) | Plasma wavelength (nm) | Average transmission ($\lambda$ =0.4–0.7μm) | Average reflection ($\lambda$ =3-25μm) | Average reflection ($\lambda$ =0.7–3μm) | Film thickness (nm) |
|---|---|---|---|---|---|---|---|---|
| 2 | 6.1 | 12.6 | 6.36 | 1678 | 86.5 | --- | 31.7 | 170 |
| 4 | 5 | 10.8 | 6.64 | 1642 | 85.6 | 61.2 | 38.0 | 167 |
| 5 | 4 | 11 | 7.50 | 1544 | 85.5 | 80.7 | 42.2 | 168 |
| 6 | 3 | 11.2 | 9.42 | 1378 | 85.2 | 89.5 | 53.4 | 168 |

Figure 3 and table 2 show the effect of oxygen flow on transmission and optical reflection of those films prepared in 400°C.

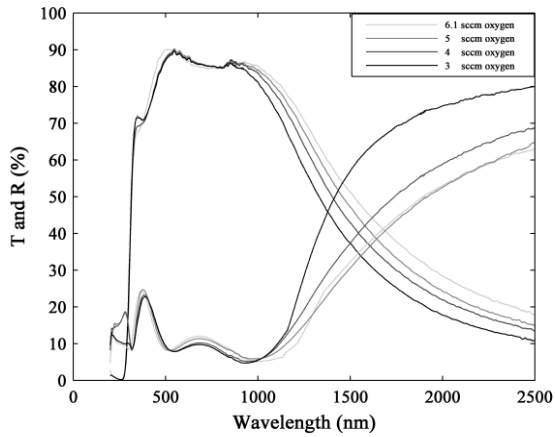

Figure 3: The effect of oxygen flow on the optical transmission and reflection for the films grown at 400°C.

Transmission doesn't change so much by oxygen flow but reflection changes severely in IR area which is due to high increase of carrier concentration because plasma wavelength moves toward shorter wavelengths by the increase of carrier concentration. Therefore, reflection begins from shorter wavelengths.

Figure 4 shows reflection in far IR area ($\lambda$ =3–25 $\mu m$) for samples No. 5 and 6. Sample No. 5 has an average reflection of 80.7% and sample No. 6 has an average reflection of 89.5% which is a good coating for manufacturing Low-E glasses. It should be noted that high IR reflectivity implies low thermal emissivity ($\varepsilon$) as a consequence of Kirchhoff's low [3].

$$\varepsilon = 1-(1+2\varepsilon_{\circ} cR_{sh})^{-2} \qquad (5)$$

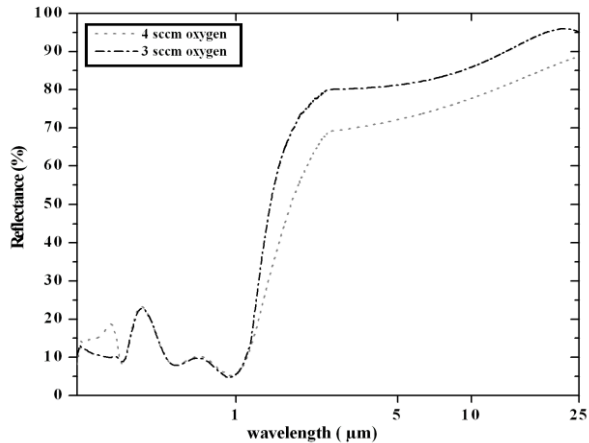

Figure 4: Reflection spectra of Low-E glass.

# Conclusion

ITO films have deposited on large glass substrate. Dependence of sputtering power and oxygen flow on electrical and optical properties has been investigated. The sheet resistance decreased by increasing the power. The Sheet resistance reached the minimum amount of 12.6Ω/sq at power of 4.3KW. Resistance increased by the increase of power. Oxygen flow is among those important parameters which can make much change in optical properties.

The most transmission and least resistance were obtained by a sputtering power of 4.3KW and an oxygen flow of 5sccm. Maximum transmission for ITO thin film with a surface resistance of 10.8Ω/sq in a wavelength of $\lambda$ =550nm has become equal to 90.31% (deposition temperature of 400°). Maximum reflection was obtained for sample No. 6 with an oxygen flow of 3sccm. In this sample, maximum average reflection was obtained in λ=3-25μm equal to 89.5% and emissivity equal to $\varepsilon = 0.11$ with a sheet resistance of 11.2Ω/sq. Transmission in the visible wavelength region of this sample in $\lambda$ =550nm is equal to 89.65%. High reflection in sample No. 6 is due to the more carrier concentration.

By analyzing the obtained data and by following deposition conditions presented in this work, film No. 6 with a sheet resistance of about 11.2Ω/sq in the sputtering condition of 4.3KW, a substrate temperature of 400° and an oxygen flow of 3sccm, optimum Low-E films made of ITO may be prepared easily on the glass substrate by using the available systems.

# References

[1] D.R. Uhlman, T. Surtwala, K. Davisdon, J.M. Boulton, G. Teowee, J. Non-Cryst. Solids 218 (1997) 113

[2] H.L. Hartnagel, A.L. Dawar, A.K. Jain, C. Jagadish, Semiconducting Transparent Thin Films, Institute of Physics Publishing, Bristol, 1995.

[3] M. Reidinger, M. Rydzek, C. Scherdel, M. Arduini-Schuster, J. Manara, Thin Solid Films 517 (2009) 3096

[4] Shuiping Huang, Zhanshan Wang, Jian Xu, Daxue Lu, Tongsuo Yuan, Thin Solid Films 516 (2008) 3179

[5] E. Hammarberg, A. Roos, Thin Solid Films 442 (2003) 222

[6] M. Bender, W. Seelig, C. Daube, H. Frankenberger, B. Ocker, J. Stollenwerk, Thin Solid Films 326 (1998) 72

[7] H. Kim, C.M. Gilmore, J. Appl. Phys. 86 (11) (1999) 6451

[8] I. Hamberg, C.G. Granqvist, J. Appl. Phys 60 (11) (1986) R123

[9] L.R. Cruz, C. Legnani, I.G. Matoso, C.L. Ferreira, H.R. Moutinho, Mater. Res. Bull. 39 (2004) 993

[10] Klaus Elmer, J. Appl. Phys. 33 (2000) R32

[11] H. Hoffmann, J.B. Webb, D.F. Williams, Appl. Phys. 16 (1978) 239

[12] A. Solieman, M.A. Aegerter, Thin Solid Films, 502 (2006)205-211

[13] J. Ederth, P. Heszler, A. Hultaker, G.A. Niklasson, C.G. Granqvist, Thin Solid Films 445 (2003)199-206